**Comment on "Optical Response of Gas-Phase Atoms at Less than $\lambda$/80 from a Dielectric Surface**" published by  K. A. Whittaker, J. Keaveney, I. G. Hughes, A. Sargsyan, D. Sarkisyan, C. S. Adams in Phys. Rev. Lett. Lett **112** 253201 (2014)

**Daniel BLOCH[1,2]**

(1) CNRS, UMR 7538, 99 Avenue J-.B. Clément, F-93430 Villetaneuse, France

(2) Laboratoire de Physique des Lasers, Université Paris 13, Sorbonne Paris-Cité,

99 Avenue J-.B. Clément, F-93430 Villetaneuse, France

e-mail: daniel.bloch@univ-paris13.fr

In a recent letter [1], Whittaker *et al.* observe the effect of the van der Waals (vW) atom-surface attraction in nanocell spectroscopy. They combine fluorescence detection, known for its high sensitivity (absence of a background for a resonance line) with a sub-wavelength thickness to observe the vW-induced red-shift and lineshape asymmetry. They claim that the method offers the ability to probe the $r^{-3}$ distance-dependence of the vW attraction to a distance as small as 11 nm from the window, through an analysis of the remote wings of the spectrum. In this comment, I show this claim to be unconvincing, notably because their analysis ignores the transient behavior of the resonant response of moving atoms, which explore the spatially-dependent potential. This transient response enhances the relative contribution of slower atoms, but the velocity distribution, notably for slow atoms, is questionable in a nanocell.

In a nanocell, spectroscopic signals integrate the contributions of atoms with various positions and velocities. Importantly, the atoms moving in the vW potential experience a resonant interaction with light in a velocity-dependent *transient* regime. For a fluorescence detection, complexity is even higher than for absorption spectroscopy in a nanocell (ref.33 of [1]) because one has to take into account the delay (implying a change in atomic position) between the atomic excitation and the fluorescence emission. In addition, the fluorescent decay is interrupted when the atom collides the wall if a non-radiative deexcitation takes place. Claiming that "*fluorescence at any given laser frequency detuning maps directly onto the distance of an atom from (...) surfaces*" (caption of fig.4 in [1]) ignores all these effects related to atomic motion. In this mapping, the authors bin the vW interaction with a step resolution as small as 1 nm (fig. 4 of [1]). This is needed because of the rapid variation of the vW potential for small distances to the surface. However, this narrow binning neglects the spectral broadening associated to the limited transit time. This broadening is considerable for any atom with a thermal velocity. "Slow" atoms (*i.e.* velocity nearly parallel to the windows) may feel more efficiently the vW interaction, but are not numerous enough to systematically dominate the response. In addition, the existence of atoms with a slow (normal) velocity (below 1 or few m/s), easily assumed for an isotropic Maxwell distribution, becomes problematic for a nanocell. This is partly because of microscopic planarity defects or windows roughness (making "parallel" velocity meaningless), and also because of the atom-surface force itself (see *e.g.* estimates in ref. 33 of [1]).

In the data processing, the fitting "Lorentzian" (*e.g.* the 86 MHz width in fig. 3 of [1]) does not relate to any predictable quantity, and solely results from an oversimplified description of spectroscopy in a nanocell (see ref. 39 of [1]). Measurements are claimed to yield a distance law $r^{-\alpha}$ with $\alpha = 3.02 \pm 0.06$. Such an accuracy is tantalizing as it largely exceeds the one currently mentioned in previous works (see ref. 21 of [1] for a 0,7µm-3µm wall-separation, ref. 33 of [1] for a 40-130 nm wall separation, or [2] for a 6-10µm distance in the Casimir-Polder regime) although they benefited from spatially resolved methods. Here, the distance range is unmentioned, and the accuracy is unreliable because consistency when varying the cell thickness and systematic effects is not assessed, nor sensitivity to residual fields [3]. An effective check of the $r^{-3}$ vW dependence remains highly desirable for fundamental theory, as alterations are predicted (see *e.g.* [4], and notably fig 5). Comparing the measured value of $C_3$ - actually $C_3(r)$- with predictions, accurate in the principle for elementary alkali-metal atom / sapphire systems, is also of a high interest. Despite the claim for a good agreement with theory, the theoretical values provided in [1], based upon current literature, are plagued by an applicability restricted to the ground state, or by a fortuitous error (for Cs, the value 2 kHz.µm$^3$ is for $D_2$ line, the published one for $D_1$ line [5] is 1 kHz.µm$^3$). Conversely, a disagreement with theory cannot be attributed to differing interactions for the various hyperfine components (see *e.g.* refs 13, 15, 33 of [1], and refs. therein).

**Acknowledgements**

I thank Martial Ducloy for a useful discussion.

**References**

[1] K. A. Whittaker, J. Keaveney, I. G. Hughes, A. Sargsyan, D. Sarkisyan, C. S. Adams, "*Optical Response of Gas-Phase Atoms at Less than λ/80 from a Dielectric Surface* ", Phys. Rev. Lett. **112**, 253201 (2014)

[2] D. M. Harber, J. M. Obrecht, J. M. McGuirk, E. A. Cornell, " *Measurement of the Casimir-Polder force through center-of-mass oscillations of a Bose-Einstein condensate* ", . Phys. Rev. A **72**, 033610 (2005)

[3 J. M. McGuirk, D. M. Harber, J. M. Obrecht, E.A. Cornell, "*Alkali-metal adsorbate polarization on conducting and insulating surfaces probed with Bose-Einstein condensates*", Phys Rev. A **69**, 062905 (2004).

[4] A. O. Caride, G. L. Klimchitskaya, V. M. Mostepanenko, and S. I. Zanette, " *Dependences of the van der Waals atom-wall interaction on atomic and material properties* ", Phys. Rev. A **71**, 042901 (2005)

[5] A. Laliotis, I. Maurin, M. Fichet, D. Bloch, M. Ducloy, N. Balasanyan, A. Sarkisyan, D Sarkisyan, "*Selective reflection spectroscopy at the interface between a calcium fluoride window and Cs vapour*" Appl. Phys. B **90**, 415 (2008)